\documentclass[12pt]{iopart}

\usepackage[dvips]{graphicx}
 
\begin{document}

\title[Excitation cross section of polarized 
atoms by polarized electrons]
{A general expression for the excitation cross section of polarized 
atoms by polarized electrons}

\author{A Kupliauskien\.{e}}

\address{Institute of Theoretical Physics and Astronomy of Vilnius University,
A.Go\v{s}tauto 12, 01108 Vilnius, Lithuania}
\ead{akupl@itpa.lt}

\begin{abstract}
The general expression for 
excitation cross section of polarized atoms by  polarized 
electrons is derived by using the methods of the theory of an  atom 
adapted for polarization.  
The special cases of the general expression 
for the description of the angular distribution and alignment of 
excited atoms  in the  case of 
polarized and non-polarized atoms as well as the magnetic dichroism 
of the total ionization cross section of polarized atoms are obtained.
The cross sections and alignment parameters for the excitation of the 
autoionizing states 2p$^5$3s$^2$ $^2$P$_{3/2}$ for Na and
3p$^5$4s$^2$ $^2$P$_{3/2}$ for K are calculated in distorted wave with exchange
approximation.
\end{abstract}


\maketitle


\vspace*{ 5mm}

\noindent
\section{Introduction}

Polarized atoms and ions presents in plasma where directed flows 
of charged particles take place.
The directed movement of these 
particles is very important  in laboratory and astrophysical
plasmas resulting in the distortion of Maxwellian distribution of
electrons \cite{Kazantsev}.
The non-equilibrium population of magnetic sub-levels or the ordering
of angular momentum of atomic particles that is called a 
self-alignment is causes
the polarization of the emitted electromagnetic fluorescence radiation
that
could be used for the fusion plasma diagnostics \cite{Mandl}.
Recently, the methods of the theory of an atom were applied for the 
derivation of the general expressions describing the interaction of polarized
photons and electrons with polarized atoms and ions [3-9].
The probability (cross section) of the interaction was expressed
as multiple expansion over the multipoles (irreducible tensors) of
the state of all particles taking part in the process both in
initial and final states.
The applied approach was an alternative to the density matrix method 
\cite{Bal2000} where the density matrix elements
were expressed via multipoles or statistical tensors.
The density matrix formalism was used for the study of some special
cases of the polarization of the particles presented in the
excitation process.
These expressions were applied for the calculations of the 
alignment of autoionizing states of alkaline atoms 
excited by electrons \cite{Pangantiwar,Na,Na1}, 
positrons \cite{Pangantiwar} or other charged particles \cite{Na}
The expressions for the description of the polarization and 
angular distribution of the radiation from unpolarized atoms 
 excited by polarized electrons were derived by Bartschat {\em et al}
\cite{Bartschat} in a general case.
These expressions were applied for the calculation of the electron 
impact excitation of the 4 $^1$P$_1$ state of calcium 
\cite{Srivastava,Chauhan}.

The main task of the present work was the derivation of a general
expression 
describing the polarization state of all particles 
taking part in the excitation of polarized atoms by polarized 
electrons with the help of the method based on the theory of an atom
\cite{K2004}.
The next section of the present work is devoted to obtaining of the
general expression. Its special cases as well the calculations of 
total cross section and alignment parameters are presented in Section~3.
The inequality
$$
{\rm fine \;\;structure\;\; splitting}\gg {\rm line\;\; width}\gg
{\rm hyperfine\;\; structure\;\; splitting}
$$ 
is also assumed.
The modifications enabling to take into account hyperfine structure
splitting can be easily made \cite{K2004,krt2001}.


\section{General expression}

For the excitation of an atom $A$ in the state
$\alpha_0 J_0M_0$ by an electron $e^-$ moving with the momentum 
{\bf p}$_1$ and spin projection $m_1$
\begin{equation}
A(\alpha_0 J_0M_0)+e^-({\rm\bf p}_1 m_1)\to
A(\alpha_1 J_1M_1) +e^-({\rm\bf p}_2 m_2),
\label{eq:es1}
\end{equation}
the cross section can be written in atomic system of units as follows:
$$
\frac{d\sigma(\alpha_0J_0M_0{\rm\bf p}_1m_1\to \alpha_1J_1M_1
{\rm\bf p}_2m_2)}{ d\Omega_2}=
\frac{p_2}{(2\pi)^2p_1}
\langle \alpha_1J_1M_1 {\rm\bf p}_2m_2|V|
\alpha_0J_0M_0{\rm\bf p}_1m_1\rangle
$$
\begin{equation}
\times
\langle \alpha_1J_1M_1 {\rm\bf p}_2m_2|V|
\alpha_0J_0M_0{\rm\bf p}_1m_1\rangle^* \delta(E_0-E_1).
\label{eq:es2}
\end{equation}
Here $V$ is the operator of the electrostatic interaction between
projectile and atomic electrons, $E_0$ and $E_1$ are the energies 
of the system atom+electron in the initial and final states,
$p_i$ is the absolute value of the momentum,$p_i=\sqrt{2\varepsilon_i}$, 
$\varepsilon_i$ is the energy of the projectile electron in the
initial ($i=1$) and final ($i=2$) states, $\alpha_2J_2M_2$ and 
{\bf p}$_2m_2$ describe the states of the excited atom and
scattered electron, respectively.

The wave function of the projectile and scattered electrons can be
expressed via expansion over Spherical harmonics
$$
|{\bf p}m\rangle = 4\pi\sum_{\lambda \mu}R_\lambda(r)
Y_{\lambda \mu}(\hat{r}) Y^*_{\lambda \mu}(\hat{p})
\xi_m(\sigma)
$$
\begin{equation}
=  \sum_{\lambda\mu} 
\sqrt{4\pi(2\lambda+1)}
R_\lambda(r)C^{(\lambda)}_\mu(\hat{r}) Y^*_{\lambda \mu}(\hat{p})
\xi_m(\sigma).
\label{eq:fotoj3}
\end{equation}
Here $\xi_m(\sigma)$ is the spin orbital of an electron,
$C^{(\lambda)}_\mu(\hat{r})$ is the operator of the spherical
function \cite{Jucys_Bandzaitis}, where $\hat{r}$ denotes the polar
and azimutal angles of the spherical coordinate system,
\begin{equation}
R^*_\lambda(r) = {\rm i}^\lambda 
\exp[{\rm i}(\sigma_\lambda(p)+\delta_\lambda)]
r^{-1} P(\varepsilon \lambda|r)
\label{eq:fotoj4}
\end{equation}
is the radial orbital of the electron in a continuum state normalized
to $\delta(\varepsilon-\varepsilon')$.
For the electron moving in the field of an ion,
the asymptotic of the Hartree orbital $P(\varepsilon \lambda|r)$ is
\begin{equation}
P(\varepsilon\lambda|r\to\infty) 
\sim (\pi p)^{-1/2} \sin(pr-\lambda \pi/2
+Z_{ef}\ln(2pr)/p +\sigma_\lambda(p)+\delta_\lambda)
\label{eq:fotoj5}
\end{equation}
with Coulomb phase
\begin{equation}
\sigma_\lambda(p)=arg \; \Gamma(\lambda+1-{\rm i}(Z_{ef}-1)/p)
\label{eq:fotoj6}
\end{equation}
and effective nuclear charge $Z_{ef}=Z-N+1$.
Here $Z$ is the nuclear charge, $N$ is the number of electrons. 
In (\ref{eq:fotoj5}),  $\delta_\lambda$ is the phase arising due to 
the deviation of the self consistent field from Coulomb one.
In the case of neutral atom, the asymptotic expression of Hartree
orbital obtains the following expression:
\begin{equation}
P(\varepsilon\lambda|r\to\infty)\sim (\pi p)^{-1/2}
\sin(pr-\lambda\pi/2+\delta_\lambda).
\label{eq:es3}
\end{equation}

The substitution of (\ref{eq:fotoj3}) into (\ref{eq:es2}) leads to the 
following expression for one transition matrix element:
$$
\langle \alpha_1J_1M_1{\rm\bf p}_1m_1|H|
\alpha_0J_0M_0{\rm\bf p}_1m_1\rangle = 
\sum_{\begin{array}{c}\tilde{M}_0,\tilde{M}_1,\tilde{m}_1,
\tilde{m}_2,\\ \lambda_1,\mu_1,
\lambda_2,\mu_2\end{array}}
[(2\lambda_1+1)(2\lambda_2+1)]^{1/2}
$$
$$
\times
\langle \alpha_1J_1\tilde{M}_1
\varepsilon_1\lambda_1\mu_2\tilde{m}_1|H|
\alpha_0J_0\tilde{M}_0\varepsilon_1\lambda_1\mu_1
\tilde{m}_1\rangle
D^{J_0}_{\tilde{M}_0M_0}(\hat{J}_0)\;
D^{*J_1}_{\tilde{M}_1M_1}(\hat{J}_1)\;
D^{\lambda_1}_{\mu_1 0}(\hat{p}_1)\;
$$
\begin{equation}
\times
D^{*\lambda_2}_{\mu_20}(\hat{p}_2)\;
D^{s}_{\tilde{m}_1m_1}(\hat{s})\;
D^{*s}_{\tilde{m}_2m_2}(\hat{s}).
\label{eq:es4}
\end{equation}
Here the possibility of the registration of the orientation of 
angular momentum of all particles with respect to different
quantization axes is taken into account. 
For the evaluation of the matrix element, the wave functions of all
particles taking part in the process were transformed to the same
system of coordinates by using the transformation procedure
\begin{equation}
|j\tilde{m}\rangle=\sum_m D^j_{m\tilde{m}}(\alpha,\beta,\gamma)
|jm\rangle.
\label{eq:dfun1}
\end{equation}
Here $D^j_{m\tilde{m}}(\alpha,\beta,\gamma)$ is the Wigner rotation matrix.

The integration over orbital and 
summation over spin variables was performed with
the help of the graphical technique of the angular momentum
\cite{Jucys_Bandzaitis,Merkelis}. 
The angular momentum diagrams used for the derivation
of the general expression of the excitation cross section of 
polarized atoms by polarized electrons are represented in figures 1 and 2.
The angular part of the matrix element (\ref{eq:es4}) is represented
in diagram E$_1$ (see figure~1).
Here the rectangle with $(kk)$ indicates the orbital and spin parts of
the electrostatic interaction operator which orbital part is defined as 
follows:
$$
V=\sum_k\frac{r^k_<}{r^{k+1}_>}(C^{(k)}_1C^{(k)}_2).
$$
Here  and 1 and 2 show the atomic and projectile electrons, 
respectively.
Other rectangles in diagram E$_1$
represent the orbital and spin parts of the 
configuration and other quantum numbers.
The circles with $D$ inside are Wigner rotation matrices.
The electron was excited from the shell which orbital quantum number
is $l_0$ to the shell which orbital quantum number is $l_1$,
and $s=1/2$.

\begin{figure}[h]
\begin{center}
\includegraphics*[height=6cm,width=10cm]{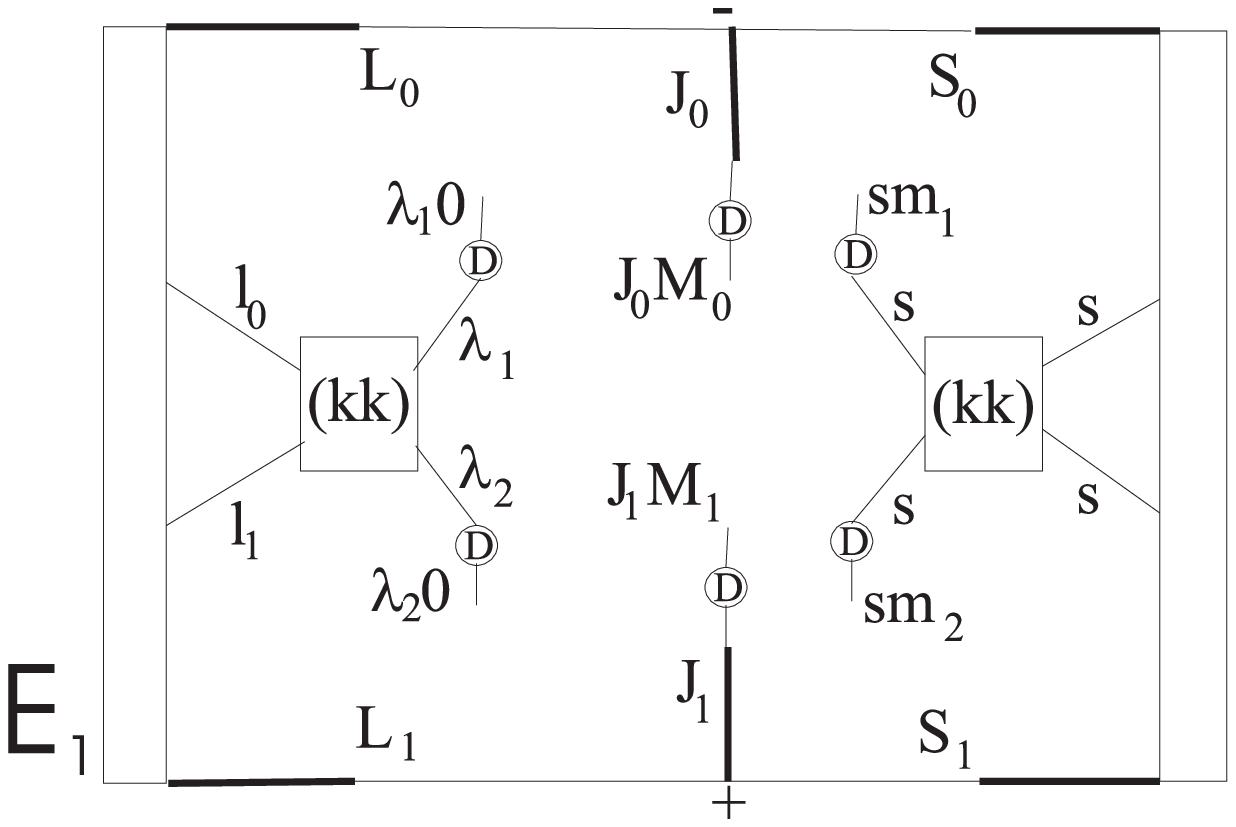}

\vspace{10mm}
\includegraphics*[height=6cm,width=10cm]{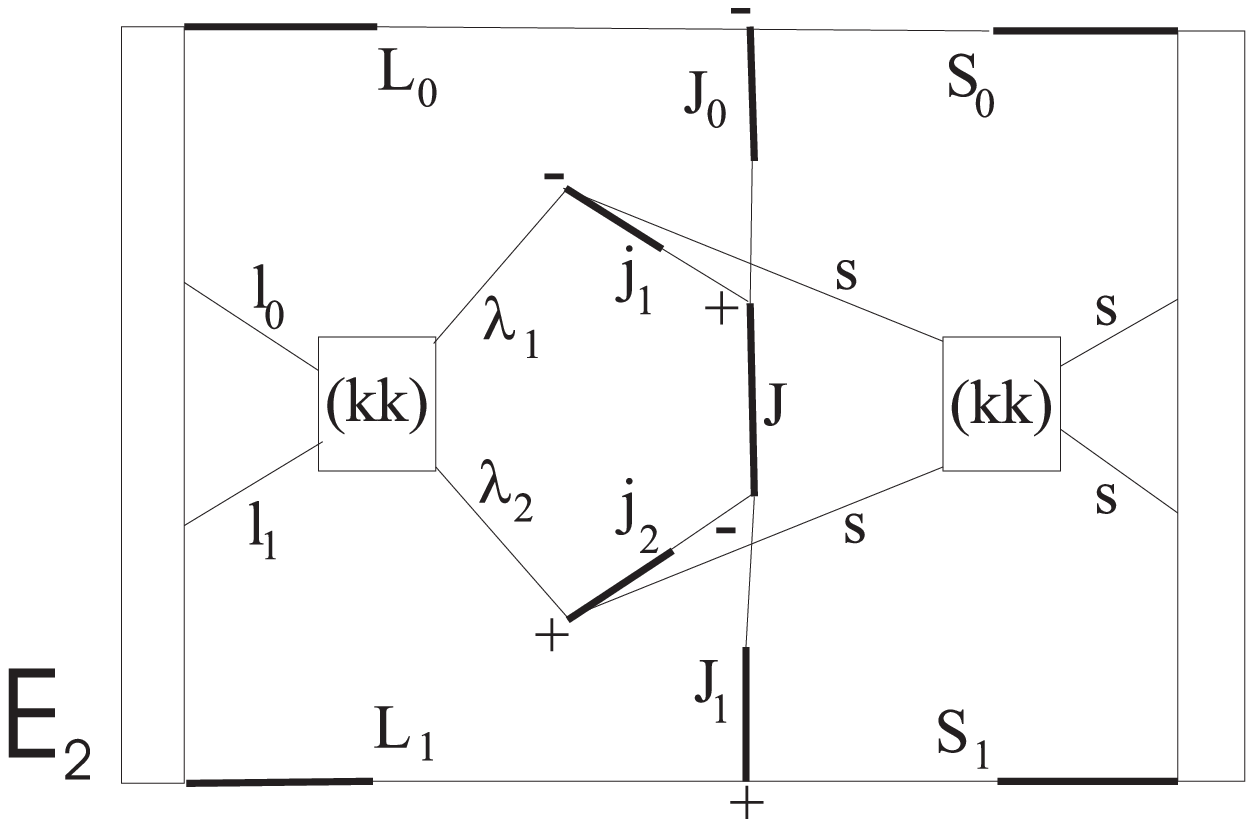}

\vspace{10mm}
\includegraphics*[height=6cm,width=9cm]{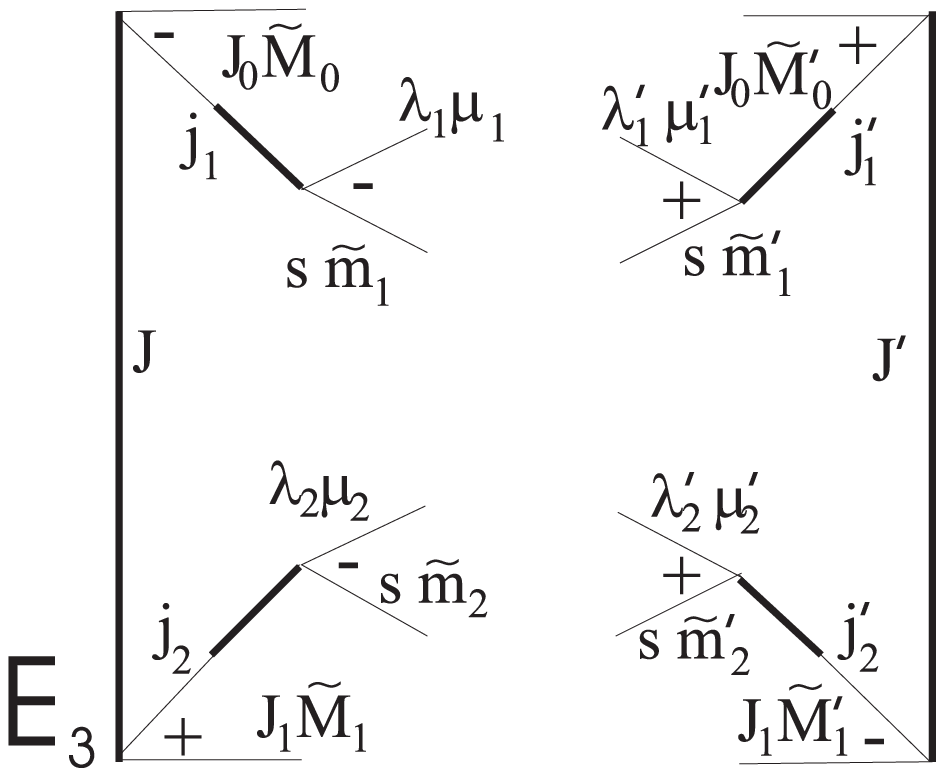}
\end{center}
{\bf Figure~1} {\small The angular momentum diagrams used for the derivation
of the general expression of the excitation cross section of 
polarized atoms by polarized electrons}.
\end{figure}


\begin{figure}[h]
\begin{center}
\includegraphics*[height=8cm,width=9cm]{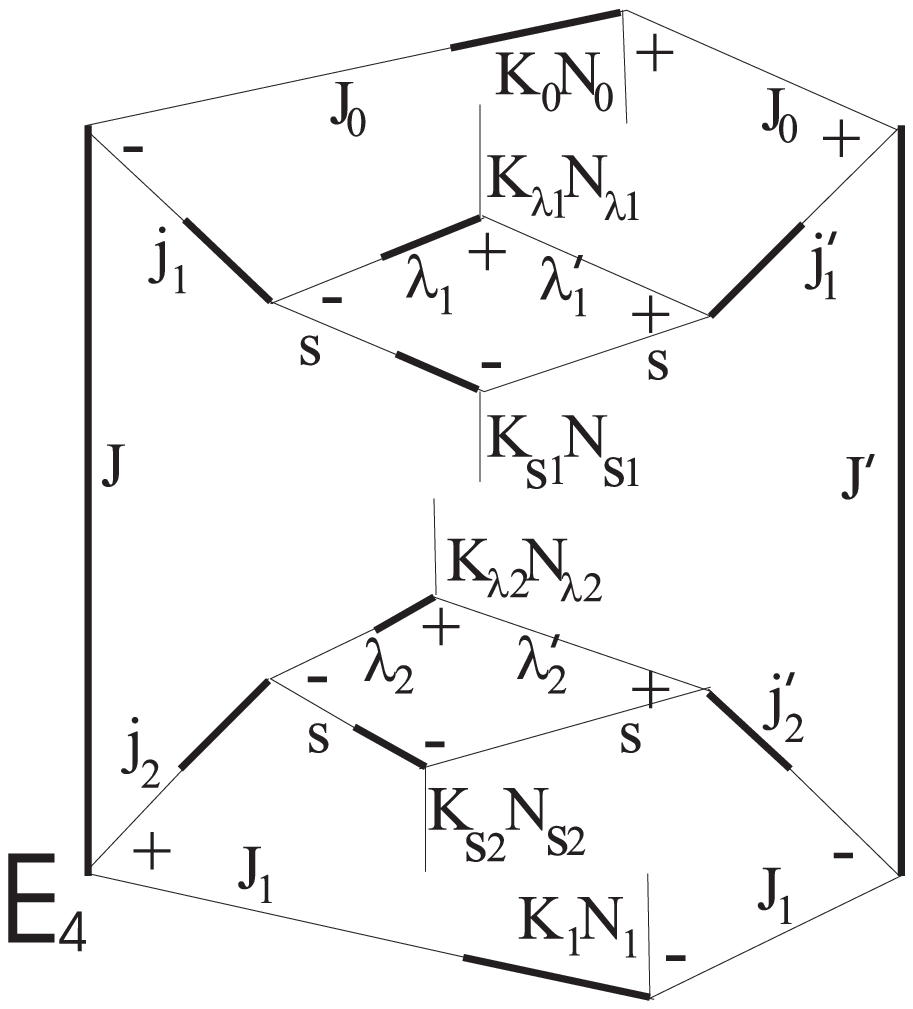}

\vspace{10mm}
\includegraphics*[height=8cm,width=8cm]{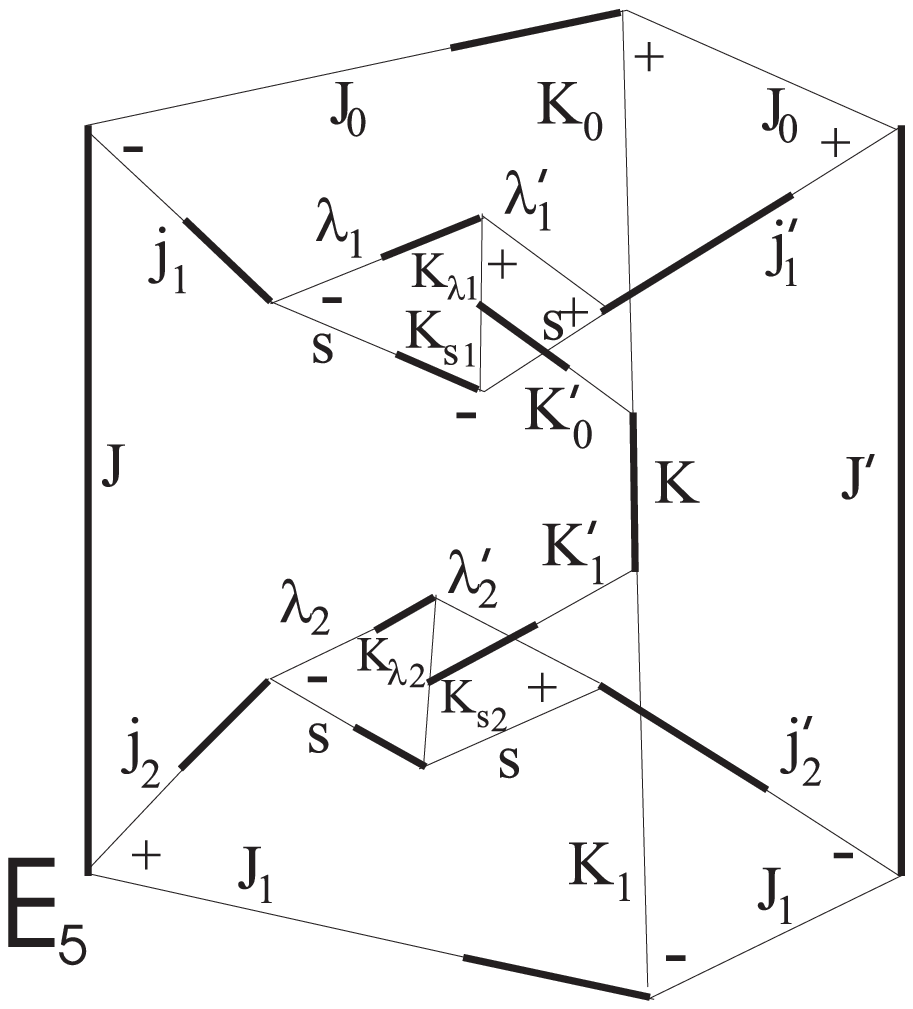}
\hspace{10mm}
\includegraphics*[height=6cm,width=4cm]{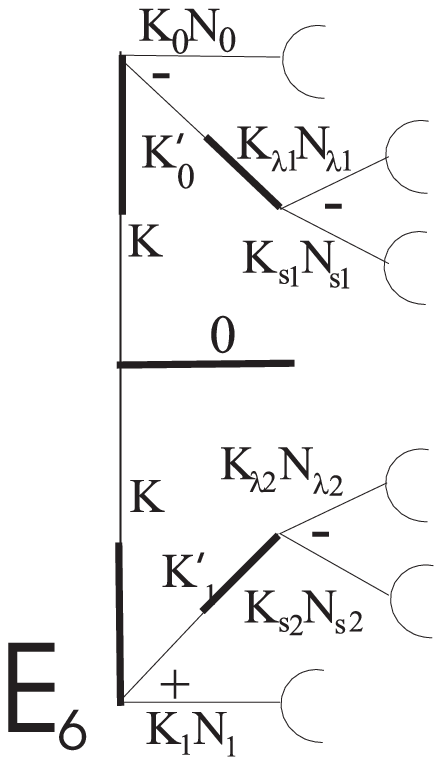}
\end{center}
{\bf Figure~2} {\small The angular momentum diagrams used for the derivation
of the general expression of the excitation cross section of 
polarized atoms by polarized electrons}.
\end{figure}

To extract the reduced matrix element from diagram E$_1$,
one needs to cut of the Wigner rotation matrices from the open lines
of each angular momentum (circles with D) and to choose the order
of the coupling of open lines.
Let us choose the following order of coupling:
$J_0,\lambda_1s(j_1)J$ and $J_1,\lambda_2s(j_2)J$.
Then the procedure of the extraction of the reduced matrix element
may be written by using the diagrams from figure~1
\begin{equation}
{\rm E}_1=\sum_{j_1,j_2,J} (2J+1) \; {\rm E}_2 \; {\rm E}'_3,
\label{eq:es5}
\end{equation}
where E$_2$ is the angular part of the diagram of reduced matrix 
element 
$\langle\alpha_1J_1,\varepsilon_2\lambda_2(j_2)J||H||\alpha_0J_0,
\varepsilon_1\lambda_1(j_1)J\rangle$
and E$'_3$ is the generalized
Clebsch-Gordan coefficient \cite{Jucys_Bandzaitis} represented by the
left side of diagram E$_3$.
The right side of diagram E$_3$ comes from the angular part
of the complex conjugate matrix element from (\ref{eq:es2}).

For further simplification of the part describing the space
rotation dependence, the following expansion can be used:
\begin{equation}
D^J_{\tilde{M}M}(\hat{J})\;D^{*J'}_{\tilde{M}'M}(\hat{J})=
\sum_{K,N}T^{*K}_{N}(\hat{J})
\left[\begin{array}{ccc}
J'&K&J\\\tilde{M}'&N&\tilde{M}
\end{array}\right].
\label{eq:es5_1}
\end{equation}
In(\ref{eq:es5_1}), the angle brackets with angular momenta inside show
the Clebsch-Gordan coefficient \cite{Jucys_Bandzaitis}, and
the tensor is defined \cite{KT2003} as
\begin{equation}
T^{*K}_{N}(\hat{J})=(-1)^{J'-M}\left[\frac{4\pi}{2J+1}\right]^{1/2}
\left[\begin{array}{ccc}
J&J'&K\\M&-M&0
\end{array}\right]
Y^*_{KN}(\theta,\phi).
\label{eq:es5_2}
\end{equation}
Six Clebsch-Gordan coefficients  
obtained by applying (\ref{eq:es5_2}) for all angular 
momenta
$\lambda_1, \lambda_2, J_0,J_1$ and two spins $s$ of free
electrons
are used to perform the summation over the projections
$\tilde{M}_0,\tilde{M}'_0,\tilde{M}_1,\tilde{M}'_1$, $\tilde{\mu}_1,
\tilde{\mu}'_1,\tilde{\mu}_1,\tilde{\mu}'_1,
\tilde{m}_1,\tilde{m}'_1,\tilde{m}_2,\tilde{m}'_2$
of the matrix element in (\ref{eq:es4}) and its complex conjugate.
This summation was performed graphically, and the result is represented
by diagram E$_4$  (see figure~2) that looks rather complicated.
But further simplifications of E$_4$ are still possible.
The closing of the open lines from diagram E$_4$ produces diagram E$_5$ 
invariant under space rotation and diagram E$_6$ describing rotation
properties of the cross section (\ref{eq:es2}).
The bow at the end of each open line stands for the tensor 
(\ref{eq:es5_2})
\cite{krt2001} in the case of $J_0,J_1$, spin $s$ and spherical
function for $\lambda_1$, and $\lambda_2$.
Four 9j-coefficients \cite{Jucys_Bandzaitis} can be obtained from 
diagram
E$_5$ by cutting it trough the lines $(J,K,J'),(j_1,K'_0,j'_1)$ and
$(j_2,K'_1,j'_2)$.

The final expression for the cross section (2) obtained by using
diagrams E$_2$, E$_5$ and E$_6$  is as follows:
$$
\frac{d\sigma(\alpha_0J_0M_0{\rm \bf p}_1 m_1 \to
\alpha_1J_1M_1 {\rm \bf p}_2m_2)} {d\Omega }
$$
$$
= 4\pi  C
\sum_{\begin{array}{c}K,K_0,K'_0,K_{\lambda 1},K_{s1},K_1\\K'_1,
K_{\lambda 2},K_{s 2}\end{array}}
{\cal B}^{\rm ex}(K_0,K'_0,K_1,K'_1,K_{\lambda 1},K_{s1},K_{\lambda 2},
K_{s 2},K)
$$
$$
\times
\sum_{\begin{array}{c}N_0,N'_0,N_{\lambda 1},N_{s1},N_1\\N'_1,
N_{\lambda 2},N_{s 2},N\end{array}}
\left[ \begin{array}{c c c}
K_{\lambda 1}&K_{s 1}&K'_0\\N_{\lambda 1}&N_{s 1}&N'_0
\end{array}\right]
\left[ \begin{array}{c c c}
K_0&K'_0&K\\N_0&N'_0&N
\end{array}\right]
\left[ \begin{array}{c c c}
K_1&K'_1&K\\N_1&N'_1&N
\end{array}\right]
$$
$$
\times
\left[ \begin{array}{c c c}
K_{\lambda 2}&K_{s 2}&K'_1\\N_{\lambda 2}&N_{s 2}&N'_1
\end{array}\right]
Y^*_{K_{\lambda 1}N_{\lambda 1}}(\hat{p}_1)\;
Y_{K_{\lambda 2}N_{\lambda 2}}(\hat{p}_2)\;
T^{*K_0}_{N_0}(J_0,J_0,M_0|\hat{J}_0)\;
T^{K_1}_{N_1}(J_1,J_1,M_1|\hat{J}_1)\;
$$
\begin{equation}
\times
T^{*K_{s1}}_{N_{s1}}(s,s,m_1|\hat{s})\;
T^{K_{s2}}_{N_{s2}}(s,s,m_2|\hat{s}),
\label{eq:es6}
\end{equation}
$$
{\cal B}^{\rm ex}(K_0,K'_0,K_1,K'_1,K_{\lambda 1},K_{s1},
K_{\lambda 2},K_{s 2},K)
$$
$$
=
\sum_{
\lambda_1,\lambda'_1,\lambda_2,\lambda'_2,j_1,j'_1,j_2,j'_2,J,J'}
(2J+1)(2J'+1)(2s+1)(-1)^{\lambda'_1+\lambda'_2}
$$
$$
\times
\langle\alpha_1J_1,\varepsilon_2\lambda_2(j_2)J||H||\alpha_0J_0,
\varepsilon_1\lambda_1(j_1)J\rangle
\langle\alpha_1J_1,\varepsilon_2\lambda'_2(j'_2)J'||H||\alpha_0J_0,
\varepsilon_1\lambda'_1(j'_1)J'\rangle^*
$$
$$
\times
[(2\lambda_1+1) (2\lambda'_1+1)(2\lambda_2+1) (2\lambda'_2+1)
(2j_1+1)(2j'_1+1)(2j_2+1)(2j'_2+1)
$$
$$
\times
(2J_0+1)(2J_1+1)(2K'_0+1)(2K'_1+1)]^{1/2}
\left[ \begin{array}{c c c}
\lambda_1&\lambda'_1&K_{\lambda 1}\\0&0&0
\end{array}\right]
\left[ \begin{array}{c c c}
\lambda_2&\lambda'_2&K_{\lambda 2}\\0&0&0
\end{array}\right]
$$
\begin{equation}
\times
\left\{ \begin{array}{c c c}
J_0&K_0&J_0\\j'_1&K'_0&j_1\\J'&K&J
\end{array}\right\}
\left\{ \begin{array}{c c c}
\lambda'_1&K_{\lambda 1}&\lambda_1\\s&K_{s1}&s\\j'_1&K'_0&j_1
\end{array}\right\}
\left\{ \begin{array}{c c c}
\lambda'_2&K_{\lambda 2}&\lambda_2\\s&K_{s2}&s\\j'_2&K'_1&j_2
\end{array}\right\}
\left\{ \begin{array}{c c c}
J_1&K_1&J_1\\j'_2&K'_1&j_2\\J'&K&J
\end{array}\right\} .
\label{eq:es7}
\end{equation}
Taking into account the equation (\ref{eq:fotoj3}) and asymptotics
(\ref{eq:fotoj5}) and (\ref{eq:fotoj6}), the constant in (\ref{eq:es2}) is
$C=4/p_1^2$.

The expression (\ref{eq:es6}) represents the most general case of the
cross section describing the excitation
 of polarized atoms by polarized electrons and enabling one
to obtain information on the angular distributions
 and spin polarization of scattered electron and the alignment of
excited atom  in non relativistic approximation.


\section{Special cases}


\subsection{Total cross section for the excitation of unpolarized
atoms}

In the case when the polarization state of excited atoms and 
scattered electrons are not registered,
the total cross section describing the excitation of unpolarized
atoms by unpolarized electrons can be obtained from the general
expression (\ref{eq:es6}) by summation over the magnetic components
of the particles in the final state and averaging over them in the
initial state as well as integration over the angles of scattered
electron.
The expression under consideration was obtained by applying the
formulas \cite{K2004}
\begin{equation}
\sum_MT^K_N(J,J,M|\hat{J})=\delta(K,0)\delta(N,0),
\label{eq:es7_1}
\end{equation}
\begin{equation}
\int^\pi_0 \int^{2\pi}_0 \sin\theta d\theta d\phi
Y_{KN}(\theta,\phi)=\sqrt{4\pi} \delta(K,0)\delta(N,0)
\label{eq:es7_2}
\end{equation}
and is as follows:
$$
\sigma(\alpha_0J_0\to \alpha_1J_1)
=\frac{1}{2(2J_0+1)}\int d\Omega \sum_{M_0,m_1,M_1,m_2}
\frac{d\sigma(\alpha_0J_0M_0{\rm \bf p}_1 m_1 \to
\alpha_1J_1M_1 {\rm \bf p}_2m_2)} {d\Omega }
$$
\begin{equation}
=
\frac{4\pi}{(2J_0+1)\varepsilon_1}{\cal B}^{\rm ex}(0,0,0,0,0,0,0,0,0)
\label{eq:es8_1}.
\end{equation}
Here
$$
{\cal B}^{\rm ex}(0,0,0,0,0,0,0,0,0)
$$
\begin{equation}
=
\sum_{\lambda_1,j_1,\lambda_2,j_2,J} (2J+1)
|\langle\alpha_1J_1,\varepsilon_2\lambda_2(j_2)J||H||\alpha_0J_0,
\varepsilon_1\lambda_1(j_1)J\rangle|^2.
\label{eq:es8_2}
\end{equation}


\subsection{The angular distribution of scattered electrons
following the excitation of unpolarized atoms}

The excitation of unpolarized atoms by unpolarized electrons is 
usual and often occurred process in  plasmas.
To obtain the expression of the differential cross section suitable
for the description of the angular distribution of scattered electrons
one needs to performer the summation  of the general expression 
(\ref{eq:es6}) over the magnetic components
of the particles in the final state and averaging over them in the
initial state.
The application of the expression (\ref{eq:es7_1}) and the choice 
of the laboratory quantization axes along the direction of the
projectile electron, 
that means $Y_{KN}(0,0)=[(2K+1)/4\pi]^{1/2}\delta(N,0)$,
leads to the expression
$$
\frac{d\sigma(\alpha_0J_0\to \alpha_1J_1{\rm\bf p}_2)}{d\Omega}
=\frac{1}{2(2J_0+1)}\sum_{M_0,m_1,M_1,m_2}
\frac{d\sigma(\alpha_0J_0M_0{\rm \bf p}_1 m_1 \to
\alpha_1J_1M_1 {\rm \bf p}_2m_2)} {d\Omega }
$$
\begin{equation}
=
\frac{\sigma(\alpha_0J_0\to \alpha_1J_1)}{4\pi}
\left[ 1 +\sum_{K>0} \beta_K P_K(\cos\theta)\right].
\label{eq:es9}
\end{equation}
Here the asymmetry parameters of the angular distribution of the
scattered electrons are defined as follows:
\begin{equation}
\beta_K=
\frac{(2K+1){\cal B}^{\rm ex}(0,K,0,K,K,0,K,0,K)}
{{\cal B}^{\rm ex}(0,0,0,0,0,0,0,0,0)}.
\label{eq:es10}
\end{equation}

In (\ref{eq:es9}), the summation parameter $K$ can acquire the
values 
$\max \{|\lambda_1-\lambda'_1|,|\lambda_2-\lambda'_2|\}
\leq K\leq \min \{\lambda_1+\lambda'_1,\lambda_2+\lambda'_2\}$
for each set of the partial wave momenta which can be very large
depending on the energy of the projectile electron.
Several partial waves are enough only for the energies close to
the excitation threshold.


\subsection{The angular distribution of scattered electrons
following the excitation of polarized atoms}

In the case of the atoms prepared in polarized state, the expression
for the differential cross section describing the angular distribution
of the scattered electrons can be also obtained by
the summation over the magnetic components
of the particles in the final state and averaging over the states 
of the spin in the initial state in the general expression 
(\ref{eq:es6}).
In the case of the choice 
of the laboratory quantization axes along the direction of the
projectile electron, the expression for the cross section is as follows:
$$
\frac{d\sigma(\alpha_0J_0M_0\to \alpha_1J_1{\rm\bf p}_2)}{d\Omega}
=\frac{1}{2}\sum_{m_1,M_1,m_2}
\frac{d\sigma(\alpha_0J_0M_0{\rm \bf p}_1 m_1 \to
\alpha_1J_1M_1 {\rm \bf p}_2m_2)} {d\Omega }
$$
$$
=
\frac{C\sqrt{4\pi}}{2}
\sum_{K_{\lambda_1},K_0,K_{\lambda_2},N_0}
[2K_{\lambda_1}+1]^{1/2}
{\cal B}^{\rm ex}(K_0,K_{\lambda_1},0,K_{\lambda 2},K_{\lambda 1},0,
K_{\lambda 2},0,K_{\lambda 2})
$$
$$
\times
\left[ \begin{array}{c c c}K_0&K_{\lambda_1}&K_{\lambda_2}\\N_0&0&N_0
\end{array}\right]
Y_{K_{\lambda_2} N_0}(\hat{p}_2)
\left[\frac{4\pi}{2J_0+1}\right]^{1/2}
(-1)^{J_0-M_0}
\left[ \begin{array}{c c c}J_0&J_0&K_0\\M_0&-M_0&0\end{array}\right]
$$
\begin{equation}
\times
Y^*_{K_0 N_0}(\hat{J}_0).
\label{eq:es11}
\end{equation}

This expression becomes more simple in the case of special
geometry of the experiment. 
If the atom is polarized along the direction of the projectile
electron,  then $N_0=0$, $M_0=J_0$, 
 $Y^*_{K_0 N_0}(0,0)=\sqrt{(2K_0+1)/4\pi}\delta(N_0,0)$,  
$Y_{K_{\lambda_2} 0}(\hat{p}_2)=\sqrt{(2K_{\lambda_2}+1)/4\pi} 
P_{K_{\lambda_2}}(\cos\theta)$, and the angle $\theta$ 
is measured from the direction of the projectile electron.
Then the expression (\ref{eq:es11}) transforms into the following
expression:
\begin{equation}
\frac{d\sigma(\alpha_0J_0M_0\!=\!J_0\to \alpha_1J_1{\rm\bf p}_2)}{d\Omega}
=
\frac{\sigma(\alpha_0J_0\to \alpha_1J_1)}{4\pi}
\left[ 1 +\sum_{K_{\lambda_2}>0} 
B_{K_{\lambda_2}} P_{K_{\lambda_2}}(\cos\theta)\right].
\label{eq:es12}
\end{equation}
Here 
$$
B_{K_{\lambda_2}}
={\cal B}^{\rm ex}(0,0,0,0,0,0,0,0,0)^{-1}
\sum_{K_0,K_{\lambda_1}}
{\cal B}^{\rm ex}(K_0,K_{\lambda_1},0,K_{\lambda 2},K_{\lambda 1},0,
K_{\lambda 2},0,K_{\lambda 2})
$$
\begin{equation}
\times
[(2J_0+1)(2K_0+1)(2K_{\lambda_1}+1)(2K_{\lambda_2}+1)]^{1/2}
\left[ \begin{array}{c c c}K_0&K_{\lambda_1}&K_{\lambda_2}\\0&0&0
\end{array}\right]
\left[ \begin{array}{c c c}J_0&J_0&K_0\\J_0&-J_0&0\end{array}\right]
\label{eq:es13}
\end{equation}
is one from the set of the asymmetry parameters of the angular
distribution of the scattered electrons.
The difference of the expression (\ref{eq:es13}) 
and that with opposite directions of {\bf J}$_0$ is equal to  
the magnetic dichroism in the electron-impact excitation
cross section of polarized atoms describing the angular distribution of the
scattered electrons.


\subsection{Magnetic dichroism in the total electron-impact 
excitation cross section  of polarized atoms}

The total cross section of the excitation of polarized atoms by
unpolarized electrons can be easily obtained by integration of
(\ref{eq:es11}) over the angles of the scattered electrons.
Then $K_{\lambda_2}=N_{\lambda_2}=0$, $K_0=K_{\lambda_2}$.
This cross section depends on the direction of the polarization of
atoms and is as follows:
$$
\sigma(\alpha_0J_0 M_0{\rm\bf p}_1\to \alpha_1J_1)
=2\pi\;C\sum_{K_0,N_0}\frac{1}{[(2J_0+1)(2K_0+1)]^{1/2}}
(-1)^{K_0+J_0-M_0}
\left[\begin{array}{c c c}J_0&J_0&K_0\\M_0&M_0&0\end{array}\right]
$$
\begin{equation}
\times
{\cal B}^{\rm ex}(K_0,K_0,0,0,K_0,0,0,0,0)
Y_{K_0N_0}(\hat{p}_1)Y^*_{K_0N_0}(\hat{J}_0).
\label{eq:es14}
\end{equation}

In the case of the choice of the quantization axis $z$ along the
direction of the projectile electron, the expression
(\ref{eq:es14}) becomes more simple:
$$
\sigma(\alpha_0J_0 M_0\to \alpha_1J_1)
=2\pi\;C\sum_{K_0}(-1)^{K_0+J_0-M_0}
\left[\frac{2K_0+1}{2J_0+1}\right]^{1/2}
\left[ \begin{array}{c c c}J_0&J_0&K_0\\J_0&-J_0&0\end{array}\right]
$$
\begin{equation}
\times
{\cal B}^{\rm ex}(K_0,K_0,0,0,K_0,0,0,0,0)
P_{K_0}(\cos\theta),
\label{eq:es15}
\end{equation}
where the angle $\theta$ of the orientation of the total angular 
momentum $J_0$ of an atom is measured from the direction of the
projectile electron.

The degree of magnetic dichroism can be defined by the parameter $a$ 
which equals to
$$
a=
\frac{\sigma(\alpha_0J_0 M_0\to \alpha_1J_1)-
\sigma(\alpha_0J_0 -\!\!M_0\to \alpha_1J_1)}
{\sigma(\alpha_0J_0 M_0\to \alpha_1J_1)+
\sigma(\alpha_0J_0 -\!\!M_0\to \alpha_1J_1)}
$$
\begin{equation}
=
\frac{\sum_{K_0=odd}B(K_0)P_{K_0}(\cos\theta)}
{\sum_{K_0=even}B(K_0)P_{K_0}(\cos\theta)},
\label{eq:es16}
\end{equation}
where
\begin{equation}
B(K)=
(-1)^{K}\sqrt{2K+1}
\left[ \begin{array}{c c c}J_0&J_0&K\\J_0&-J_0&0\end{array}\right]
{\cal B}^{\rm ex}(K,K,0,0,K,0,0,0,0).
\label{eq:es17}
\end{equation}
Here the values of the summation parameter are $K_0\leq 2J_0$.
If the total angular momentum of an atom $J_0$ is directed along
and  opposite directions of the projectile electron, then
 $M_0=J_0$, $P_{K_0}(0)=1$, and the parameter of the diamagnetic
dichroism is as follows:
\begin{equation}
a=
\frac{\sum_{K_0=odd}B(K_0)}
{\sum_{K_0=even}B(K_0)}.
\label{eq:es18}
\end{equation}

For small values of the total angular momentum of an atom, 
the expression of the magnetic dichroism is presented as an example.
In the case of $J_0=1/2$, it is as follows:
\begin{equation}
a= - \sqrt{3}
\frac{{\cal B}^{\rm ex}(1,1,0,0,1,0,0,0,0)}
{{\cal B}^{\rm ex}(0,0,0,0,0,0,0,0,0)}.
\label{eq:es19}
\end{equation}

For $J_0=1$, the magnetic dichroism is defined by the expression:
\begin{equation}
a=
\frac{-(3/\sqrt{2})\;{\cal B}^{\rm ex}(1,1,0,0,1,0,0,0,0)}
{{\cal B}^{\rm ex}(0,0,0,0,0,0,0,0,0) + \sqrt{5/2}\;
{\cal B}^{\rm ex}(2,2,0,0,2,0,0,0,0)}.
\label{eq:es20}
\end{equation}


\subsection{The alignment of unpolarized atoms
excited by electron-impact }

Usually the alignment of the excited atom effects the characteristics 
of the second step processes \cite{K2004a,KT2003}.
Therefore the special expression should be used in this case.
It can be obtained with the help of the method presented by 
Kupliauskien\.{e} \cite{K2004,K2004a} for  (\ref{eq:es6})
and is as follows:
$$
\frac{d\sigma_{K_1N_1}(\alpha_0J_0M_0{\rm \bf p}_1 m_1 \to
\alpha_1J_1 {\rm \bf p}_2m_2)} {d\Omega }
$$
$$
= 4\pi  C[2K_1+1]^{1/2}
\sum_{\begin{array}{c}K,K_0,K'_0,K_{\lambda 1},K_{s1}\\K'_1,
K_{\lambda 2},K_{s 2}\end{array}}
{\cal B}^{\rm ex}(K_0,K'_0,K_1,K'_1,K_{\lambda 1},K_{s1},K_{\lambda 2},
K_{s 2},K)
$$
$$
\times
\sum_{\begin{array}{c}N_0,N'_0,N_{\lambda 1},N_{s1}\\N'_1,
N_{\lambda 2},N_{s 2},N\end{array}}
\left[ \begin{array}{c c c}
K_{\lambda 1}&K_{s 1}&K'_0\\N_{\lambda 1}&N_{s 1}&N'_0
\end{array}\right]
\left[ \begin{array}{c c c}
K_0&K'_0&K\\N_0&N'_0&N \end{array}\right]
\left[ \begin{array}{c c c}
K_1&K'_1&K\\N_1&N'_1&N  \end{array}\right]
$$
$$
\times
\left[ \begin{array}{c c c}
K_{\lambda 2}&K_{s 2}&K'_1\\N_{\lambda 2}&N_{s 2}&N'_1
\end{array}\right]
Y^*_{K_{\lambda 1}N_{\lambda 1}}(\hat{p}_1)\;
Y_{K_{\lambda 2}N_{\lambda 2}}(\hat{p}_2)\;
T^{*K_0}_{N_0}(J_0,J_0,M_0|\hat{J}_0)\;
T^{*K_{s1}}_{N_{s1}}(s,s,m_0|\hat{s})\;
$$
\begin{equation}
\times
T^{K_{s2}}_{N_{s2}}(s,s,m_1|\hat{s}).
\label{eq:es8}
\end{equation}

The cross section describing the alignment of excited atoms 
can be obtained
by the integration over the angles of  scattered electrons, summation
over the magnetic components of the spin of scattered electrons and
averaging over the states of atoms and electrons in the initial state
of the expressions (\ref{eq:es8}).
Its expression is as follows:
$$
\sigma_{K_1N_1}(\alpha_0J_0{\rm\bf p}_1\to \alpha_1J_1)=
\sum_{K_1}
\frac{1}{2(2J_0+1)}\int d\Omega \sum_{M_0,m_1,m_2}
\frac{d\sigma_{K_1N_1}(\alpha_0J_0M_0{\rm \bf p}_1 m_1 \to
\alpha_1J_1M_1 {\rm \bf p}_2m_2)} {d\Omega }
$$
\begin{equation}
=\frac{2\pi C}{2J_0+1}[4\pi(2K_1+1)]^{1/2}
\sum_{K_1}{\cal B}^{\rm ex}(0,K_1,K_1,0,K_1,0,0,0,K_1)
Y_{K_1N_1}(\hat{p}_1).
\label{eq:es21}
\end{equation}
More simple expression can be obtained by coinciding the 
laboratory $z$ axis with  the direction of  {\bf p}$_1$:
\begin{equation}
\sigma_{K_10}(\alpha_0J_0\to \alpha_1J_1)
=\frac{2\pi C}{2J_0+1}\sum_{K_1}(2K_1+1)
{\cal B}^{\rm ex}(0,K_1,K_1,0,K_1,0,0,0,K_1).
\label{eq:es22}
\end{equation}

The alignment parameters obtains the following expression:
\begin{equation}
A_{K_1}=
\frac{(2K_1+1)
{\cal B}^{\rm ex}(0,K_1,K_1,0,K_1,0,0,0,K_1)}
{{\cal B}^{\rm ex}(0,0,0,0,0,0,0,0,0)}.
\label{eq:es23}
\end{equation}
where $K_1=2,4,...,2J_1$.
In the case of $J_1=0$ and $J=1/2$, the excited atoms can not be 
aligned because of $K_1\leq 2$.
For $J_1=1$, the alignment of excited atoms is described by the
expression:
\begin{equation}
A_2=
\frac{5{\cal B}^{\rm ex}(0,2,2,0,2,0,0,0,2)}
{{\cal B}^{\rm ex}(0,0,0,0,0,0,0,0,0)}.
\label{eq:es25}
\end{equation}


\section{Applications and discussion}

The derived expressions for the total cross section (\ref{eq:es8_1}),
reduced matrix elements 
$\langle\alpha_1J_1,\varepsilon_2\lambda_2(j_2)J||H||
\alpha_0J_0,\varepsilon_1\lambda_1(j_1)J\rangle$, the asymmetry parameter
of the angular distribution of the scattered electrons $\beta_K$
(\ref{eq:es10}), the alignment parameters $A$ (\ref{eq:es23}) of 
unpolarized atoms excited in the case of unpolarized electrons as well 
as the degree of magnetic dichroism $a$ (\ref{eq:es18}) of the excitation of
polarized atoms by unpolarized electrons are used in the creation of computer
program in FORTRAN. 
The calculations of the total cross section and alignment parameter for
the excitation the autoionizing states of 2p$^5$3s$^2$ $^2$P$_{3/2}$ for Na
and 3p$^5$4s$^2$ $^2$P$_{3/2}$ for K are carried out as an example.
The cross sections are displayed in figure~3, and the alignment 
parameters are shown in figure~4.
The distorted wave (DW) with exchange approximation and multiconfiguration
wave functions \cite{Froese} are used. 
The calculations in plane wave Born (PWB) approximation were also
performed in the present work.
The correlation effects are important and decrease the values of the cross
section about 7\% for Na and 8\% for K. 
It can be expected because the dominant expansions are 
0.96~2p$^5$3s$^2$+0.26~2p$^5$3p$^2$ and 0.95~3p$^5$4s$^2$+0.30~3p$^5$4p$^2$
$-$0.09~3p$^5$4s($^1$P)3d$-$0.09~3p$^5$4s($^3$P)3d for Na and K, respectively.
In the case of Na, the cross sections for the term $^2$P were calculated 
in order to make the comparison with experimental data \cite{Fuerstein}.
The cross sections calculated in DW and PWB
approximations are in close agreement with those obtained by
Borovik \etal  \cite{Borovik2005} in the same approximations,
therefore they are not presented in figure~3 for comparison.

\begin{figure}
\begin{center}
\includegraphics*[height=8cm,width=10cm]{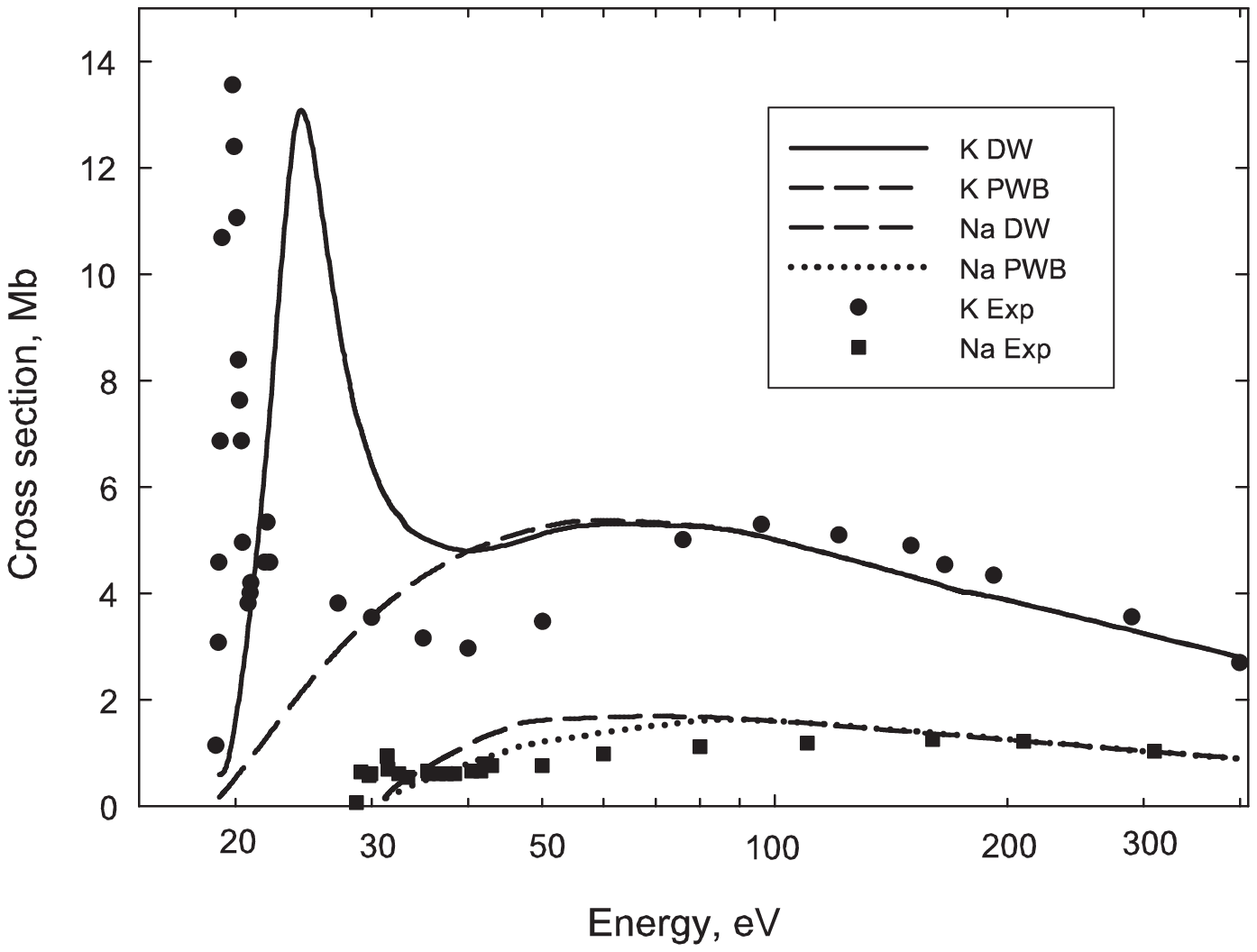}
\end{center}
{\bf Figure~3} 
{\small The comparison of calculated electron-impact excitation cross sections
2p$^6$3s$\to$2p$^5$3s$^2$ $^2$P and 
3p$^6$4s$\to$3p$^5$4s$^2$ $^2$P$_{3/2}$ with experimental data
 for Na \cite{Fuerstein} and K \cite{Borovik2005,KBBZ2006}.
}
\end{figure}

The results from figure~3 show that the values of calculated cross sections
merge with experimental data \cite{Fuerstein,Borovik2005,KBBZ2006} for the
incident electron energies about three times larger than excitation threshold
both for Na and K.
A sharp increase of the excitation cross section near threshold for K is
caused by the exchange scattering as it disappears in the cross section
calculated in DW without exchange.

\begin{figure}
\begin{center}
\includegraphics*[height=8cm,width=10cm]{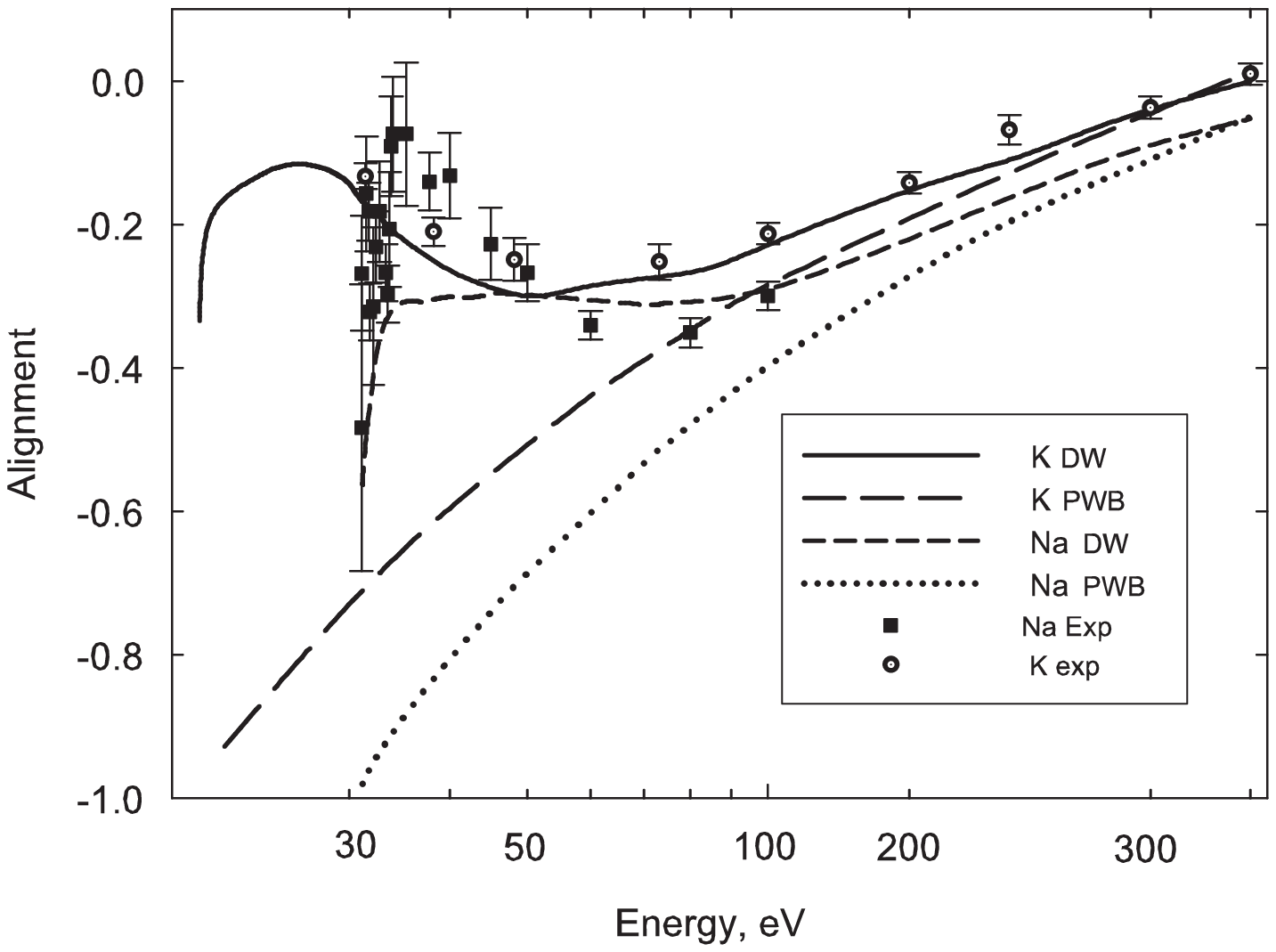}
\end{center}
{\bf Figure~4} 
{\small The alignment parameters of electron-impact excited state
2p$^6$3s$\to$2p$^5$3s$^2$ $^2$P$_{3/2}$ for Na and 
3p$^6$4s$\to$3p$^5$4s$^2$ $^2$P$_{3/2}$ for K calculated in DW and PWB
approximations.
Experimental data are presented for Na \cite{Na1} and  K \cite{Materstock}.
 }
\end{figure}

The results of figure~4 show that calculated in DW approximation alignment 
parameters of the state 3p$^5$4s$^2$ $^2$P$_{3/2}$ in K excited by
electron impact are in good agreement with experimentally determined 
\cite{Materstock} over a broad range of incident electron energies.
This agreement can be explained that an important part of the excitation
mechanisms cancels in the ratio of the excitation cross sections for
different magnetic sublevels.
The values of the alignment parameter for K of the present work are close
to those obtained by Materstock \etal \cite{Materstock} in DW  approximation
 with exchange using some optical potential.
For the excitation energies exceeding the threshold twice, the calculated
alignment parameters for Na in the state 2p$^5$3s$^2$ $^2$P$_{3/2}$ are also
in good agreement with experimental data \cite{Na1} and 
values calculated by applying {\em R}-matrix approach \cite{Na1}.
Significant deviations of the alignment parameters
calculated in the present work from those of experimental data
and {\em R}-matrix calculations can be noticed close to the excitation 
threshold indicating the importance of the correlations in the continuum.
For large excitation energies, the PWB approximation can be used to calculate
the alignment parameters for Na and K atoms.


\section{Concluding remarks}

The general expression for the cross section describing the excitation of
polarized atoms by polarized electrons as a multiple expansion over the
multipoles of the states of all particles participating in the process
both in the initial and final states was obtained for the first time.
A simple way to derive expressions for special cases by using the general
expression is described.
In the case of the excitation of unpolarized atoms by unpolarized 
electrons, more simple expressions for the total cross section, parameters 
of the alignment of excited atoms and the asymmetry parameters of the
angular distribution of the scattered electrons are obtained from the
general expression.
The derivation of the degree of magnetic dichroism as well as the parameters
of the asymmetry of the angular distribution of scattered electrons in the 
case of the excitation of polarized atoms by unpolarized electrons are also 
demonstrated by using the general expression.
The special expressions are implemented into computer code. 
The calculations of the total cross section and alignment parameter for
the excitation the autoionizing states of 2p$^5$3s$^2$ $^2$P$_{3/2}$ for Na
and 3p$^5$4s$^2$ $^2$P$_{3/2}$ for K are carried out as an example.
 The calculated in DW approximation alignment 
parameters of the state 3p$^5$4s$^2$ $^2$P$_{3/2}$ in K excited by
electron impact are in good agreement with those experimentally determined 
over a broad range of incident electron energies.
Significant deviations of the alignment parameters calculated in 
the present work from those of experimental data
and  {\em R}-matrix calculations  noticed close to the excitation threshold
indicate the importance of the correlations in the continuum.

\ack
The work was partially funded by the Joint Taiwan-Baltic Research
project and Ministry of Education and Science of Lithuania No. SUT-901
and  by the European Commission, project RI026715, Baltic Grid.

\end{document}